\documentclass[aps,pre,onecolumn,showpacs,showkeys,superscriptaddress,nobalancelastpage,nofootinbib,floatfix,longbibliography]{revtex4-2}
\usepackage{graphics}
\usepackage{graphicx}
\usepackage{dcolumn}
\usepackage{bm}
\usepackage{comment} 
\usepackage{xcolor}
\usepackage{bookmark}
\usepackage{tensor}
\usepackage{amsmath}
\usepackage{amssymb}
\usepackage{dsfont}
\usepackage{float}
\usepackage[utf8]{inputenc}
\usepackage{romannum}
\usepackage{moreverb}
\usepackage{subfigure}
\usepackage{hyperref}
\usepackage[capitalize]{cleveref}

\begin{document}
\pagenumbering{arabic}
\newcommand{\ket}[1]{\left| #1 \right\rangle}
\newcommand{\bra}[1]{\left\langle #1 \right|}
\newcommand{\braket}[1]{\left\langle #1 \right\rangle}
\newcommand{\bket}[1]{\langle #1 \rangle}
\newcommand{\beq}{\begin{equation}}
\newcommand{\eeq}{\end{equation}}
\newcommand{\bea}{\begin{align}}
\newcommand{\eea}{\end{align}}
\newcommand{\bq}{\begin{quote}}
\newcommand{\eq}{\end{quote}}
\newcommand{\oem}{\color{blue}}
\newcommand{\blk}{\color{black}}
 \newtheorem{thm}{Theorem}
 \newtheorem{cor}[thm]{Corollary}
 \newtheorem{lem}[thm]{Lemma}
 \newtheorem{prop}[thm]{Proposition}
 \newtheorem{defn}[thm]{Definition}
 \newtheorem{rem}[thm]{Remark}

\pagestyle{plain}
\title{\bf Nonlocality enhanced precision in quantum polarimetry via entangled photons}
\author{ Ali Pedram}
\email{apedram19@ku.edu.tr}
\affiliation{Department of Physics, Koç University, Istanbul, Sarıyer 34450, Türkiye}
\author{ Vira R. Besaga}
\email{vira.besaga@uni-jena.de}
\affiliation{Institute of Applied Physics, Abbe Center of Photonics, Friedrich-Schiller-Universität Jena, 07745 Jena, Germany}
\author{ Frank Setzpfandt}
\email{f.setzpfandt@uni-jena.de}
\affiliation{Institute of Applied Physics, Abbe Center of Photonics, Friedrich-Schiller-Universität Jena, 07745 Jena, Germany}
\affiliation{Fraunhofer Institute for Applied Optics and Precision Engineering IOF, 07745 Jena, Germany}
\author{Özgür E. Müstecaplıoğlu}
\email{omustecap@ku.edu.tr}
\affiliation{Department of Physics, Koç University, Istanbul, Sarıyer 34450, Türkiye}
\affiliation{TÜBİTAK Research Institute for Fundamental Sciences, 41470 Gebze, Türkiye}
\affiliation{Faculty of Engineering and Natural Sciences, Sabancı University, Tuzla 34956, Türkiye}
\pagebreak

\begin{abstract}
A nonlocal quantum approach is presented to polarimetry, leveraging the phenomenon of entanglement in photon pairs to enhance the precision in sample property determination. By employing two distinct channels, one containing the sample of interest and the other serving as a reference, the conditions are explored under which the inherent correlation between entangled photons can increase measurement sensitivity. Specifically, we calculate the quantum Fisher information (QFI) and compare the accuracy and sensitivity for the cases of single sample channel versus two channel quantum state tomography measurements. The theoretical results are verified by experimental analysis. The theoretical and experimental framework demonstrates that the nonlocal strategy enables enhanced precision and accuracy in extracting information about sample characteristics more than the local measurements. Depending on the chosen estimators and noise channels present, theoretical and experimental results show that noise-induced bias decreases the precision for the estimated parameter. Such a quantum-enhanced nonlocal polarimetry holds promise for advancing diverse fields including material science, biomedical imaging, and remote sensing, via high-precision measurements through quantum entanglement.
\end{abstract}
\maketitle


\section{Introduction}
\label{sec:intro}
The concept of light polarization is fundamental, depicting the directional behavior of electric field oscillations within the electromagnetic context. In cases where light encompasses diverse polarization states, it remains unpolarized. Polarizers, materials adept at transforming unpolarized light into a single polarization, play a vital role in this context. Meanwhile, analyzing how light's polarization evolves upon interaction with materials offers insights into their inherent properties, forming the core of polarimetry~\cite{Vitkin:2015,He:2021}.

In the realm of contemporary metrology, the pursuit of refined techniques for enhancing precision measurements is unceasing. One intriguing avenue is quantum enhanced metrology in which the goal is to use quantum resources such as entanglement and squeezing, for transcending classical precision bounds~\cite{PhysRevD.23.1693,MA201189,PhysRevLett.118.140401,doi:10.1126/science.1138007,PhysRevLett.112.103604,PhysRevLett.96.010401,RevModPhys.90.035005,doi:10.1126/science.1104149}. Entangled photon pairs, generated through processes like spontaneous parametric down-conversion~\cite{Burnham:1970, Hong:1985}, have enabled remarkable applications in quantum communication and cryptography~\cite{PhysRevLett.82.2594,PhysRevLett.70.1895,Bouwmeester1997,doi:10.1126/science.282.5389.706,PhysRevA.65.032302,PhysRevA.68.042317,Kimble2008,Luo2023}. Due to this fact, quantum entanglement, a phenomenon where particles achieve intrinsic correlation regardless of spatial separation, has captured particular attention in optical quantum metrology and it is revealed that prudent use of entanglement can be effective in mitigating classical noise, potentially elevating precision beyond classical limits~\cite{doi:10.1080/00107510802091298,PhysRevLett.105.013603,PhysRevA.80.052114,PhysRevLett.94.020502,Wolfgramm2013,10.1103/PhysRevLett.113.250801,PhysRevA.92.062303,PhysRevLett.114.110506,10.1103/PhysRevA.94.012339,PhysRevA.94.012101,PhysRevA.97.042112,PhysRevLett.129.070502}. This leads to the natural question, whether the integration of quantum entangled light could potentially amplify the precision of polarimetric measurements.

To reveal the sensitivity of quantum states of light for estimating physical parameters of optical samples one needs to treat light-matter interaction in a fully quantum manner. To this end attempts have been made to describe the polarization states within the framework of quantum optics~\cite{LUIS2016283,PhysRevA.48.4702,10.1103/PhysRevLett.80.1202,PhysRevA.71.023810,Soderholm_2012,Goldberg:21}. It is well established that the classical definition for degree of polarization fails to capture polarization characteristics of quantum states of light and quantum states can contain so called "hidden polarization" or higher order polarization to which higher moments of the Stokes vectors contribute. Several works have been published to demonstrate the discrepancy between quantum and classical concepts of polarization and also new measures for quantifying the quantum degree of polarization have been proposed~\cite{KLYSHKO1992349,doi:10.1080/09500349908231279,PhysRevA.66.013806,PhysRevA.72.033813,LUIS2007173,PhysRevLett.105.153602,BJORK20104440,PhysRevA.75.053806,PhysRevA.84.045804,10.1103/PhysRevA.87.043821,PhysRevA.87.043814,Sanchez-Soto_2013}.

One can describe polarization transformation of light interacting with a medium using Jones calculus or more generally using Mueller calculus~\cite{hecht2017optics}. Interaction of the optical signal with a medium can induce various effects on the signal including attenuation due to absorption or reflection, diattenuation of the signal due to dichroism, rotation of its Stokes vector due to birefringence or depolarization due to random scattering. In the classical theory of polarimetry, in case that the Mueller matrix is not singular, the Lu-Chipman decomposition (polar decomposition) can be utilized to effectively establish an input-output relation between input and output Stokes vectors decomposing the original Mueller matrix into three elementary matrices with well defined optical properties and parameterized by their respective physical parameters (rotation angles, depolarization factors, etc.)~\cite{Lu:1996}. It is important to note that there has been attempts to generalize the framework of classical Mueller polarimetry and ellipsometry into the quantum domain~\cite{PhysRevResearch.2.023038,Rudnicki:20}. Finding the ultimate measurement sensitivity of these parameters using optimal quantum probes and measurements is of utmost importance. It is shown that NOON states and anti-coherent states can saturate the Heisenberg limit for rotation sensing~\cite{PhysRevA.78.052333,PhysRevA.98.032113,Martin2020optimaldetectionof,ZGoldberg_2021,Goldberg2023beyondtranscoherent,PhysRevApplied.20.024052}. The problem of rotation sensing is also studied considering different polar decompositions with optical depolarization and diattenuation noise channels~\cite{Pedram_2024}. Furthermore, Fock and two-mode squeezed vacuum (TMSV) states are optimal for sensing loss channels and outperform coherent states for estimating the loss parameter in the bosonic channels, however they can't beat the shot noise limit~\cite{PhysRevA.79.040305,PhysRevLett.112.120405,PhysRevA.77.053807,PhysRevA.89.023845,Hayat:99,PhysRevLett.59.2555,PhysRevA.104.052615,JAKEMAN1986219,Losero2018,PhysRevLett.98.160401,PhysRevLett.121.230801,PhysRevApplied.16.044031}. For a depolarization channel it is shown that Heisenberg limit scaling is not possible, however using correlated probes and ancilla-assisted schemes increases the estimation accuracy~\cite{PhysRevA.66.022308,PhysRevA.92.032324,Frey_2011,PhysRevA.72.052334}. In general, it was argued that no parameter which is encoded on the density matrix by a non-unitary operator or a convex combination of unitary operators can be estimated beyond the shot-noise limit~\cite{4655455}. However, later research revealed that controlled and error-corrected metrology and environmental monitoring can be utilized to restore the Heisenberg limit for certain problems~\cite{PhysRevLett.112.150802,Zhou2018,Albarelli2018restoringheisenberg,PhysRevA.101.032347}. Recently it was demonstrated that using an entangled photonic source, one can non-destructively probe single-layer cell cultures and differentiate between different optical elements in the setup with a fewer number of projective measurements,~\cite{zhang2024probing,besaga2023nonlocal}. However, quantum advantage in terms of metrology is not studied in detail these works. It has been shown theoretically and experimentally that using entangled photons one can achieve a higher precision in ellipsometric measurements~\cite{Abouraddy:02,PhysRevA.70.023801}. Additionally, ghost polarimetry using polarization entangled photons and utilizing correlations between the Stokes operators in the reference and sample beams have gained attraction in recent years because of increased sensitivity in parameter estimation, specifically for the cases in which having a few number of photons is advantageous in order not to perturb the material that is being probed~\cite{Hannonen:20,Magnitskiy:20,Magnitskiy:22,PhysRevA.106.062601,Magnitskiy2021}. 

\begin{figure}[htbp] 
\centering
\includegraphics[width=0.5\textwidth]{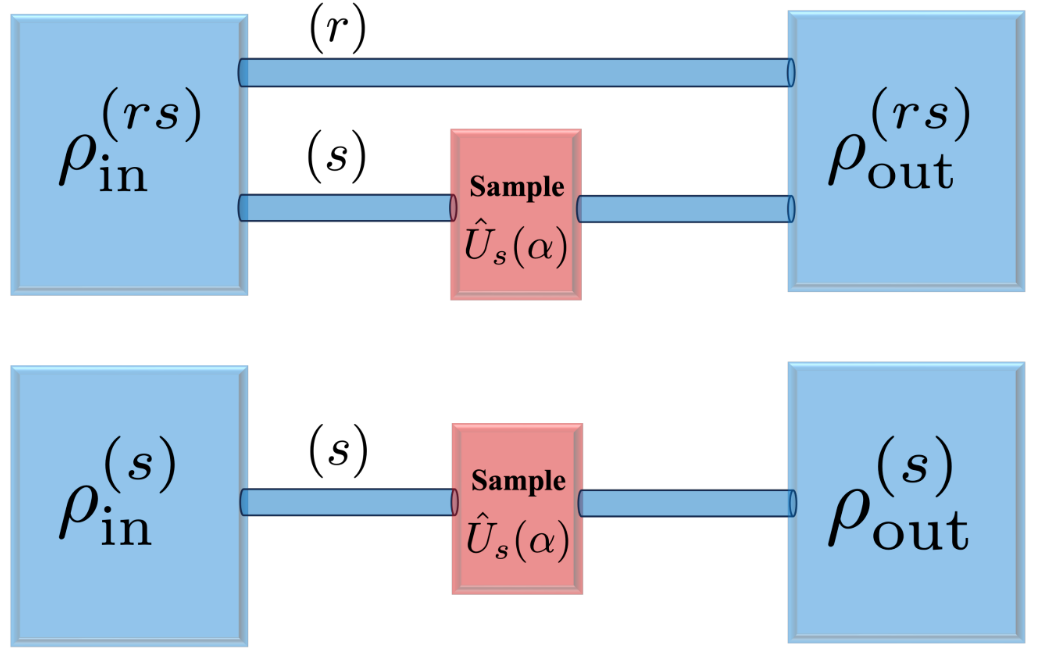} 
\caption{((Top Panel) Nonlocal Quantum Polarimetry Using Entangled Photons. In this setup, a pair of photons initially prepared in a polarization-entangled state $\rho_{\text{in}}^{(rs)}$ is directed through two distinct spatial channels. One of these paths incorporates a sample denoted as $(s)$, while the other path serves as the reference denoted as $(r)$. The sample channel can induce changes in the polarization state of the photon, a phenomenon described by a single-parameter operation $\hat U_s(\alpha)$. On the contrary, the $(r)$ channel applies an identity operation, $\hat I_r$, to the photon it contains. By employing quantum state tomography (QST) and conducting measurements in various polarization bases, the resulting output state $\rho_{\text{out}}^{(rs)}$ is reconstructed. (Bottom Panel) Local Quantum Polarimetry Using Single Photons. This scenario involves a single photon that is prepared in the reduced state of the two-photon entangled state, specifically $\rho_{\text{in}}^{(s)}=\text{Tr}_r(\rho_{\text{in}}^{(rs)})$. Subsequently, this photon traverses the sample channel. Employing QST allows for the reconstruction of the output state.}
\label{fig:setup}
\end{figure}

Within the framework of our quantum polarimetry approach, we endeavor to harness the unique nonlocal attributes inherent to entangled photons. Illustrated in Fig.~\ref{fig:setup}, our approach employs two separate pathways: a channel containing the sample (s) and a reference channel (r), with each accommodating one photon and the photons in the two channels being entangled. By capitalizing on the nonlocal correlations between these photons, we aim to discern subtle polarization changes with enhanced precision. This nonlocal methodology has the potential to mitigate local noise and disturbances that traditionally challenge single-channel polarimetry methods. Through comprehensive theoretical analysis, we examine the viability of the nonlocal approach, particularly by evaluating the QFI, in comparison to the single-channel approach. Hereby, our work contributes to the understanding of fundamental nonlocal quantum phenomena in metrology and proposes a practical avenue for utilizing nonlocal quantum states in polarimetry. This avenue could potentially impact high-sensitivity measurements across scientific disciplines.

		
\section{Preliminaries: Local and Nonlocal Polarimetry} \label{sec:Preliminaries}


We consider an entangled pair of photons, signal and idler, generated by spontaneous parametric down conversion (SPDC). They are prepared in a Bell-type entangled state in the canonical polarization basis. We assume the  initial state
can be ideally prepared as a pure state such that
\beq
\label{eq:initialState}
\ket{\psi_{\text{in}}}=\frac{1}{\sqrt{2}}(\ket{HV}+\ket{VH}),
\eeq
where $H$ and $V$ refers to horizontal and vertical polarizations, respectively.
For the sake of brevity, we adopt a convenient notation  $\ket{AB}=\ket{A}_r\otimes\ket{B}_s$ with $A,B\in\{H,V\}$ such that the first letter corresponds to the photon's state in the reference channel, while the second letter pertains to the sample channel's photon state.

A more explicit connection to the Bell state can be established by delving further into the polarization states. These states can be effectively expressed in terms of Fock number states, wherein the quantities $n_H$ and $n_V$ denote the counts of horizontally ($H$) and vertically ($V$) polarized photons. Specifically, $\ket{H} = \ket{n_H=1,n_V=0} \equiv \ket{1_H}$ and $\ket{V} = \ket{n_H=0,n_V=1} \equiv \ket{1_V}$. The introduction of a vacuum state encompassing both polarizations, denoted as $\ket{\text{vac}} := \ket{0,0} \equiv \ket{0}$, facilitates the introduction of creation and annihilation operators corresponding to $H$ and $V$ polarized photons. This can be mathematically articulated as follows:
\begin{equation}
\ket{n_k} = \frac{(\hat a_k^\dag)^{n_k}}{\sqrt{n_k!}}\ket{0},
\end{equation}
where $k=H,V$. The translation of polarization mode number states into spin states $\ket{S,S_z}$ is achieved through the utilization of the total spin, defined as $S = (n_H + n_V)/2$, and the $S_z=(n_H-n_V)/2$-component of spin (or magnetic quantum number). This translation assumes that the $z$-axis is chosen as the quantization axis. Consequently, the horizontal polarization state, $\ket{H} = \ket{n_H=1,n_V=0} = \ket{S=1/2,S_z = 1/2} \equiv \ket{\uparrow}$, can be identified as the spin-up state of a spin-$1/2$ particle. Similarly, we can equate $\ket{V}$ with the spin-down state, $\ket{\downarrow}$.
In light of this, Equation (\ref{eq:initialState}) can be interpreted as a Bell state, associated with the entanglement of two spin-$1/2$ particles; although the photons are not spin-$1/2$ particles.  For transverse waves, away from radiation sources in the far field, we can thus use an effective spin-$1/2$ representation for the polarization states of a single photon.

In the context of a single photon polarization and spin-$1/2$ analogy, the utilization of polarization mode creation and annihilation operators as Schwinger boson representation of spin operators allows for the identification of generators corresponding to the SU(2) spin algebra. These generators serve as observables that capture the quantum polarization attributes of photons. Named after their resemblance to classical Stokes parameters, these observables are referred to as Stokes operators. In the context of canonical polarizations, they are expressed as follows:
\begin{align}
\label{eq:StokesOps}
\hat S_0 &= \frac{1}{2}(\hat a_H^\dag a_H + a_V^\dag a_V), \\
\hat S_1 &= \frac{1}{2}(\hat a_H^\dag a_V + a_H a_V^\dag), \\
\hat S_2 &= -\frac{i}{2}(a_H^\dag a_V - \hat a_V^\dag a_H), \\
\hat S_3 &= \frac{1}{2}(\hat a_H^\dag a_H - a_V^\dag a_V),
\end{align}
where $\hat S=(\hat S_1 \hat S_2 \hat S_3)^T$ constitutes the quantum Stokes vector. Notably, this vector commutes with $\hat S_0$, while its individual components adhere to the spin algebra: $[\hat S_i,\hat S_j] = i\epsilon_{ijk}\hat S_k$. Within the domain of a single photon state, the Stokes operators serve as comprehensive descriptors of the photon's density matrix, thus enabling quantum state tomography (QST) of the single photon state through measurement of these operators. Consequently, any alterations in quantum polarization attributes resulting from photon-sample interactions manifest as changes in the expectation values of the Stokes operators, denoted as $S_k = \text{Tr}[\rho\hat S_k]$.

The Bell state arising from Eq.~(\ref{eq:initialState}) embodies a spin triplet state $\ket{S=1,S_z=0}$, emerging from the amalgamation of two spin-$1/2$ entities. This insight prompts the proposal of extending the concept of single photon Stokes parameters to scenarios involving two or more photons, thereby encapsulating the polarization attributes of composite multiphoton systems. We remark that multiphoton nonlocal generalizations of Stokes parameters are different than generalized Stokes operators written for spin-1 atomic Bose Einstein condensate to describe their magnetization properties or for photons that can have longitudinal polarization components in near or intermediate radiation zones, where transverve wave approximation fails. Such local generalizations of Stokes operators uses generators of SU(3) algebra, introduced by Gell-Mann to explore strong interaction in particle physics~\cite{Griffiths:2008}. In contradistinction to the "local" Stokes operators linked with individual photons, this "nonlocal" extension of Stokes parameters introduces a heightened level of complexity in terms of visualization and interpretation. Given the equivalence between Stokes parameter measurements and the direct determination of density matrices, multiphoton Stokes parameters can be equated with the process of quantum state tomography (QST) applied to a multiphoton density matrix \cite{James:2001}. Remarkably, the density matrix itself offers a means to quantify the degree of polarization. This is achieved by assessing the Q function across angular coordinates, integrating it across the entire angular domain, and subsequently gauging the deviation from the Q function of unpolarized photons \cite{PhysRevA.66.013806}. Furthermore, the density matrix of a single photon can be decomposed using Stokes operators, which intriguingly coincide with the decomposition obtained via Pauli matrices. This property can be further generalized to encompass multiphoton scenarios. 

$N$-photon generalized nonlocal Stokes operators are proposed by direct generalization of single photon Stokes operators~\cite{Goldberg:21}, which are equivalent to Pauli spin-$1/2$ operators $\hat S_i=\hat\sigma_i$ with $i=0,1,2,3$ to  
\begin{equation}\label{stokes_pauli_expansion}
  \hat S_{i_1...i_N}=\hat \sigma_{i_1}\otimes ... \otimes \hat \sigma_{i_N}.
\end{equation}
Here, $\hat \sigma_0 = \hat I$ is the identity operator, and we have $\text{Tr}(\hat\sigma_i\sigma_j)=2\delta_{ij}$. Simlar to the Pauli operator decomposition of single photon density matrix, the $N$-photon density matrix can be written as
\begin{equation}\label{dm_stokes_expansion}
\hat{\rho}=\frac{1}{2^N}\sum_{i_1=0}^3 .. .\sum_{i_N=0}^3 S_{i_1...i_N} \hat S_{i_1...i_N},
\end{equation}
where $S_{i_1...i_N}=\text{Tr}(\rho \hat S_{i_1...i_N})$ are the multiphoton nonlocal Stokes parameters.
 
In our case, we have $N=2$ photons in an initially entangled state. One photon will interact with the sample in the $s$-channel while the other channel is empty and taken as the reference ($r$-channel). Accordingly, we determine the output state by evaluating the map
\beq
\ket{\psi_{\text{out}}}=\left(\hat I_r\otimes \hat U_s(\alpha) \right)\ket{\psi_{\text{in}}}.
\eeq
In this formulation, we account for the scenario wherein the sample under consideration could either take the form of a linear polarizer (LP) or a quarter wave plate (QWP), each characterized by a distinct angular parameter denoted as $\alpha$. In the case of the LP, $\alpha$ signifies the transmission angle relative to the vertical axis within the laboratory frame. Conversely, for the QWP, $\alpha$ denotes the angle defining the orientation of the fast axis of the birefringent crystal concerning the vertical axis within the laboratory frame.

The effect of a QWP in the canonical polarization basis can be described by~\cite{Barnett2009}
\beq \label{eq:QWP}
\hat U_{\text{QWP}}(\alpha)= \begin{bmatrix}
	\cos^2\alpha +i\sin^2\alpha& (i-1)\sin\alpha\cos\alpha  \\
	 (i-1)\sin\alpha\cos\alpha & 	i\cos^2\alpha +\sin^2\alpha
\end{bmatrix},
\eeq
while for an LP we have~\cite{Barnett2009}
\beq \label{eq:LP}
\hat U_{\text{LP}}(\alpha)= \begin{bmatrix}
	\sin^2\alpha& \sin\alpha\cos\alpha  \\
	\sin\alpha\cos\alpha & 	\cos^2\alpha
\end{bmatrix}.
\eeq

In the forthcoming section, our analysis is dedicated to the evaluation of the output state, which arises from a predetermined initial entangled state and distinct models characterizing the sample. The density matrix components of this output state are ascertained via QST. These elements, when combined in specific ways, give the  nonlocal $2$-photon Stokes parameters. Our objective is to examine  the potential manifestation of a quantum advantage that may arise due to quantum nonlocality. More specifically, we endeavor to discern whether the adoption of this nonlocal quantum approach offers any discernible advantages over conventional local measurements, which are confined to single channels. After finding the output states, we evaluate the QFI to answer this question.


\section{Evaluation of the output state } \label{sec:OutputState}


The determination of the output state follows a straightforward matrix product procedure. For the specific case of a linear polarizer (LP) sample, the outcome can be succinctly expressed as:
\begin{equation}\label{out_2plp}
\ket{\psi_{\text{out}}}=A(\ket{HH}+\ket{VV})+C\ket{HV}+B\ket{VH},
\end{equation}
Here, the coefficients are assigned as $A=\sin\alpha\cos\alpha$, $B=\sin^2\alpha$, and $C=1-B =\cos^2\alpha$.

Discrepancies between the theoretical predictions and experimental outcomes for the output state can be attributed to a couple of distinct factors. Firstly, the experimental initial state possesses slight deviations from the ideal entangled state, necessitating its description through a density matrix $\rho_{\text{in}} $, accounting for these imperfections. Consequently, these deviations propagate to the output state, warranting a depiction as the density matrix $\rho_{\text{out}} $. Secondly, real-world sample conditions diverge from the ideal LP or QWP, introducing diattenuation and depolarization influences on the photon's polarization state. These alterations can be effectively modeled through supplementary operations alongside $\hat U_s(\alpha)$, formalized using Mueller matrices in accordance with the Lu-Chipman decomposition \cite{Lu:1996}, extended to the quantum domain~\cite{PhysRevResearch.2.023038}.

In the present analysis, the modest impact of these effects, coupled with the close proximity of theoretical and experimental outcomes, has led us to omit a detailed exploration of imperfections in the results. Nevertheless, it is noteworthy that such imperfections could potentially enrich the polarimetric analysis of an unknown sample. The constellation of parameters encompassing retardation angle, diattenuation, and depolarization characteristics can serve as distinctive signatures for materials, offering a means of characterization and classification vis-à-vis datasets from other samples.

Further comparison between the theoretical output state and experimental findings has consistently yielded nearly perfect agreement  in the case of the QWP as well. An essential divergence between the LP and QWP scenarios lies in the emergence of imaginary components within the density matrix elements in the latter case.

The input state for the local, single $s$-channel scenario is given by the reduced state
\beq
\rho_{\text{in}}^{(s)}=\frac{1}{2}(\ket H_s\bra H_s+\ket V_s\bra V_s),
\eeq
which is maximally mixed, unpolarized, state with $S_0 = 1/2, S_1=S_2=S_3=0$. After the LP,
the output state becomes
 \begin{equation}\label{out_1plp}
\rho_{\text{out}}^{(s)}=B\ket H_s\bra H_s+C\ket V_s\bra V_s+A(\ket H_s\bra V_s+\ket V_s\bra H_s).
\end{equation}
The Stokes parameters of the output state are given by $S_0 = 1/2, S_1 = A, S_2 = 0, S_3 = B-1/2$,
which shows that LP can polarize the initially unbpolarized photon.  

Following a similar procedure we find the output of the two photon entangled state after being acted upon by QWP to be
\begin{equation}\label{out_2pqwp}
  \ket{\psi_{\text{out}}}=\frac{(i-1)A}{\sqrt{2}}(\ket{HH}+\ket{VV})+\frac{iC+B}{\sqrt{2}}\ket{HV}+\frac{C+iB}{\sqrt{2}}\ket{VH}.
\end{equation}

For the case of local polarimetry, the state at the output of QWP becomes
 \begin{equation}\label{out_1pqwp}
  \rho_{\text{out}}^{(s)}= \frac{1}{2}(\ket H_s\bra H_s+\ket V_s\bra V_s).
\end{equation}

It is evident that the utilization of both single-channel local and two-channel non-local quantum polarimetry, encompassing the examination of local and nonlocal Stokes parameters, or equivalently, the elements of local or nonlocal states, furnishes a viable means to ascertain the unknown parameters associated with the sample, represented in this context by the angle $\alpha$.
However, which strategy proves more precise in this pursuit? In order to address this question rigorously, we proceed to evaluate the QFI in the subsequent section.


\section{ Comparison of the precision  } \label{sec:FisherInfo}

QFI quantifies the maximum achievable information about a parameter using a parameterized density matrix. The Quantum Cramer-Rao bound (QCRB) establishes a lower bound on the variance of any unbiased estimator $\hat{\alpha}$ of the parameter $\alpha$ such that for a single repetition of the measurement of $\theta$, the variance of the estimator is lower bounded by the reciprocal of the QFI~\cite{10.1116/1.5119961}. The mathematical expression of the QCRB for an unbiased estimator $\hat{\alpha}$ is
\begin{equation}\label{qcrb}
  \text{Var}(\alpha)\geq\frac{1}{mF_Q},
\end{equation}
in which $m$ is the number of observations and $F_Q$ is the QFI. For a pure state QFI can be expressed as~\cite{10.1116/1.5119961}
\begin{equation}\label{qfi_pure}
F_Q = 4[\langle\partial_\alpha\psi|\partial_\alpha\psi\rangle
-|\langle\psi|\partial_\alpha\psi\rangle|^2].
\end{equation}
For a mixed state the QFI becomes~\cite{10.1116/1.5119961}
\begin{equation}\label{qfi_mixed}
  \mathcal{F}=\sum_i\frac{(\partial_\alpha\lambda_i)^2}{\lambda_i}+
2\sum_{i\neq j}\frac{(\lambda_i-\lambda_j)^2}{\lambda_i+\lambda_j}|\langle e_i|\partial_\alpha e_j\rangle|^2.
\end{equation}
We aim to find the estimation precision of the parameter $\alpha$ using QFI for both local and non-local scenarios. For the case of the LP, the output state for the single channel has the matrix form
\beq
\rho_{\text{out}}=\begin{bmatrix}
B & A \\
A & 1-B
\end{bmatrix},
\eeq
whose eigenvalues are $\lambda_0 = 0$ and $\lambda_1=1$. The corresponding eigenvectors can be written as $e_0 = (\cos\alpha -\sin\alpha)^T$ and $e_1 = (\sin\alpha \cos\alpha)^T$. For the case of an entangled initial state and two-channel nonlocal polarimetry, the output state is a pure state as given in~\cref{out_2plp}. Substituting them into the quantum Fisher information formula gives
\begin{equation}\label{qfi_lp}
\begin{aligned}
  \mathcal{F}_{\text{LP}}^{\text{L}} = 4, \\
  \mathcal{F}_{\text{LP}}^{\text{NL}} = 8.
\end{aligned}
\end{equation}
Both expressions are independent of $\alpha$. Hence we find $\mathcal{F}_{\text{LP}}^{\text{NL}}=2\mathcal{F}_{\text{LP}}^{\text{L}}$. Similar calculations for local metrology for the case of QWP gives 
\begin{equation}\label{qfi_qwp}
\begin{aligned}
  \mathcal{F}_{\text{QWP}}^{\text{L}} = 0, \\
  \mathcal{F}_{\text{QWP}}^{\text{NL}} = 8.
\end{aligned}
\end{equation}
It is clear that the output state for the local case is parameter independent as given in~\cref{out_1pqwp} and we can't use it to give any estimate of $\alpha$. Therefore the non-local approach is far superior than the local one for the case of QWP. It is instructive to check the QFI for equally weighted superposition states of vertical and horizontal polarization modes for the two photon polarimetry scheme. Using such a state for two-photon LP and QWP settings we get 
\begin{equation}\label{qfi_sp}
\begin{gathered}
  \mathcal{F}_{\text{LP}}^{\text{SP}} = 4, \\
  \mathcal{F}_{\text{QWP}}^{\text{SP}} = 8-4\text{cos}^2(2\alpha).
\end{gathered}
\end{equation}
From~\cref{qfi_sp} we see that for local polarimetry, the QFI for the superposition state $(|H\rangle + |V\rangle)/\sqrt{2}$ for the case of LP is equivalent to that of completely mixed state given in~\cref{qfi_lp}. For the case of QWP, upon using the superposition state as a probe, one gains an overall increase in efficiency compared with the completely mixed probe states. However, we see that the QFI becomes dependent on $\alpha$ in this case and it takes values between 4 and 8. We see that polarimetry for QWP and LP with entangled photons not only yields a larger result for QFI, but entanglement is also necessary to gain information independent of the specific choice of $\alpha$. 


\section{Quantum State Tomography} \label{sec:qst}

QST is the process of reconstructing a quantum state from the measurement outcomes of experiments on an ensemble of identical quantum states~\cite{James:2001,Altepeter:2004,dariano2003quantum,ALTEPETER2005105}. Considering that the measurements are represented by the positive operator-valued measurements (POVM) $\hat{\Pi}_i=\ket{\psi_i}\bra{\psi_i}$, we know that measuring $\hat{\Pi}_i$ will produce an outcome with label $i$ whose probability is calculated using Born rule
\begin{equation}\label{born_rule}
p_i =\text{Tr}[\hat{\rho}\hat{\Pi}_i].
\end{equation}
Experimentally, one measures the POVM, $\mathcal{N}$ times using $\mathcal{N}$ copies of the unknown state $\hat{\rho}$. The average number of coincidence counts will be given by
\begin{equation}\label{coin_count}
  n_i = \mathcal{N}\text{Tr}[\hat{\Pi}_i\hat{\rho}].
\end{equation}
Therefore in QST one approximates $p_i$ by $f_i=n_i/\mathcal{N}$. For $\mathcal{N}\rightarrow\infty$ the probability for these measurement outcomes is given by $p_i = n_i/\mathcal{N}$. Assuming that the Hilbert space is finite dimensional and taking ${\hat{E}_i}$ as a set of orthonormal Hermitian basis of traceless operators, we can express the density matrix as:
\begin{equation}\label{dm_general_expansion}
  \hat{\rho} = \frac{\hat{I}}{d}+\sum_{i=1}^{d^2-1}r_i\hat{E}_i.
\end{equation} 
Here, $r_i$ are real numbers and $d$ is the dimensionality of the Hilbert space. Substituting~\cref{dm_general_expansion} in~\cref{born_rule} and approximating the probabilities with frequencies we get.
\begin{equation}\label{dm_general_expansion}
  f_i \approx \frac{\text{Tr}[\hat{\Pi}_i]}{d}+\sum_{i=1}^{d^2-1}r_i\text{Tr}[\hat{\Pi}_i\hat{E}_i].
\end{equation} 
Assuming that $\tilde{f}_i=f_i-\text{Tr}[\hat{\Pi}_i]{d}$ and $\tilde{\Pi}_{i,j}=\text{Tr}[\hat{\Pi}_i\hat{E}_i]$ we can write this equation in a matrix form as $\bm{\tilde{f}}\approx\bm{\tilde{\Pi}r}$. The goal is to find the vector of expansion coefficients $\bm{r}$ using $\bm{\tilde{f}}$ which is only possible if the matrix $\bm{\tilde{\Pi}}$ is invertible. This condition is satisfied for informationally complete sets of POVMs that span the space of density matrices acting on the Hilbert space.

In what follows in this section, we will present a basic review of optical polarization QST for a single and two qubit systems. In polarization QST the experimental data is a collection of coincidence counts of the photodetectors implementing a collection of projective measurements. In polarization measurements, the usual states used for projective measurements are horizontal ($\ket{H}$) and vertical ($\ket{V}$) linear polarized states, right ($\ket{R}$) and left ($\ket{L}$) circular states and diagonal ($\ket{D}$) and anti-diagonal ($\ket{A}$) polarization states. Hence, the goal is to estimate the state $\hat{\rho}$ based on the information of the coincidence counts $n_i$ for each projective measurement $\ket{\psi_i}\bra{\psi_i}$. For a single qubit~\cref{dm_stokes_expansion} can be written as
\begin{equation}\label{q1_qst}
  \hat{\rho} = \frac{1}{2}\sum_{i=0}^{3}\frac{S_i}{S_0}\sigma_i.
\end{equation}
Writing the Stokes operators in terms of the POVMs one can find their expectation values as
\begin{equation}\label{q1_stokes}
\begin{gathered}
\mathcal{N} = n_H+n_V = n_D+n_A = n_R+n_L,\\
S_1 = \frac{1}{\mathcal{N}} (n_D-n_A),\\
S_2 = \frac{1}{\mathcal{N}} (n_R-n_L),\\
S_3 = \frac{1}{\mathcal{N}} (n_H-n_V).
\end{gathered}
\end{equation}
Upon substituting these equations into~\cref{q1_qst} and assuming $S_0=1$ (no photon loss) one can estimate the one qubit density matrix using the coincidence numbers of the six POVMs. For two qubits~\cref{dm_stokes_expansion} can be written as
\begin{equation}\label{q2_qst}
  \hat{\rho} = \sum_{i=1}^{16}r_i\hat{\Gamma_i},
\end{equation}
in which the basis $\hat{\Gamma_i}$ are the two-qubit Stokes operators. In~\cref{q2_qst} $r_i = \text{Tr}[\hat{\Gamma_i}\hat{\rho}]$ are the expansion coefficients of the density matrix for their respective two-qubit Stokes operators given as a $16\times1$ vector. The coincidence counts and the expansion coefficients are connected via
\begin{equation}\label{count_coeff}
  n_i = \mathcal{N} \sum_{i=1}^{16}(B^{-1})_{i,j}n_j.
\end{equation} 
The $16\times16$ matrix $B$ is defined as $B_{i,j} = \bra{\psi_i}\Gamma_j\ket{\psi_i}$. It is shown that the density matrix can be expanded as
\begin{equation}\label{dm2_counts}
  \hat{\rho} = \frac{1}{\mathcal{N}}\sum_{i=1}^{16}\hat{M}_i n_i,
\end{equation}
in which the matrix $\hat{M}_i$ are $4\times4$ matrices defined via
\begin{equation}\label{m_matrix}
  \hat{M}_i = \sum_{i=1}^{16}(B^{-1})_{i,j}\Gamma_j.
\end{equation}
To partially mitigate the detrimental effects of experimental systematic errors on the reconstructed density matrices, one can utilize maximum likelihood estimation (MLE). In MLE, the objective is to explore the entire space of permissible, physically valid density matrices in order to identify the one that maximizes the likelihood of yielding the observed measurement outcomes. During this optimization process, to ensure that the resulting states are physical, the density matrices are written as their normalized Cholesky decomposition
\begin{equation}\label{dm_cholesky}
  \hat{\rho}(\vec{t}) = \frac{\hat{T}^{\dagger}\hat{T}}{\text{Tr}[\hat{T}^{\dagger}\hat{T}]}.
\end{equation}
By construction,~\cref{dm_cholesky} yields a positive, Hermitian and normalized density matrix. The operator $\hat{T}$ is a triangular matrix and is parametrized as 
\begin{equation}\label{tmat_cholesky}
  \hat{T}(\vec{t}) = 
\begin{bmatrix}
  t_1 & 0 & \cdots & 0 \\
  t_{2^n+1}+it_{2^n+1} & t_2 & \cdots &0 \\
  \cdots & \cdots & \cdots & 0 \\
  t_{4^n-1}+it_{4^n} & t_{4^n-3}+it_{4^n-2} & t_{4^n-5}+it_{4^n-4} & t_{2^n}.
\end{bmatrix}
\end{equation}
The distribution of coincidence numbers is of Poisson type but for large coincidence counts, due to the central limit theorem, one can approximate this using a normal distribution. For a normally distributed coincidence counts, the likelihood that the parametrized density matrix~\cref{dm_cholesky} produces the measured counts is
\begin{equation}\label{likelihood}
P(\vec{n},\vec{t})=\frac{1}{N_{norm}}\prod_{i}\text{exp}\left[\frac{(\mathcal{N}\text{Tr}[\hat{\Pi}_i\hat{\rho}(\vec{t})]-n_i)^2}{2\mathcal{N}\text{Tr}[\hat{\Pi}_i\hat{\rho}(\vec{t})]}\right].
\end{equation}
The task is to maximize this likelihood function by finding the optimal values for the parameters in $\vec{t}$. To simplify the procedure, one can minimize the negative of the log-likelihood function instead of maximizing the likelihood. Doing so, the function that MLE minimizes becomes
\begin{equation}\label{liglikelihood}
\mathcal{L}(\vec{n},\vec{t})=\sum_{i}\text{exp}\left[-\frac{(\mathcal{N}\text{Tr}[\hat{\Pi}_i\hat{\rho}(\vec{t})]-n_i)^2}{2\mathcal{N}\text{Tr}[\hat{\Pi}_i\hat{\rho}(\vec{t})]}\right].
\end{equation}
Hence, through the process of MLE quantum state tomography, due to~\cref{dm_cholesky} one is guaranteed to get a valid density matrix which minimizes the penalty function~\cref{liglikelihood} and therefore is most likely to produce the experimentally measured coincidence counts.


\section{Experimental Implementation} \label{sec:Experiment}


\begin{figure}[ht] 
\centering
\includegraphics[width=.85\textwidth]{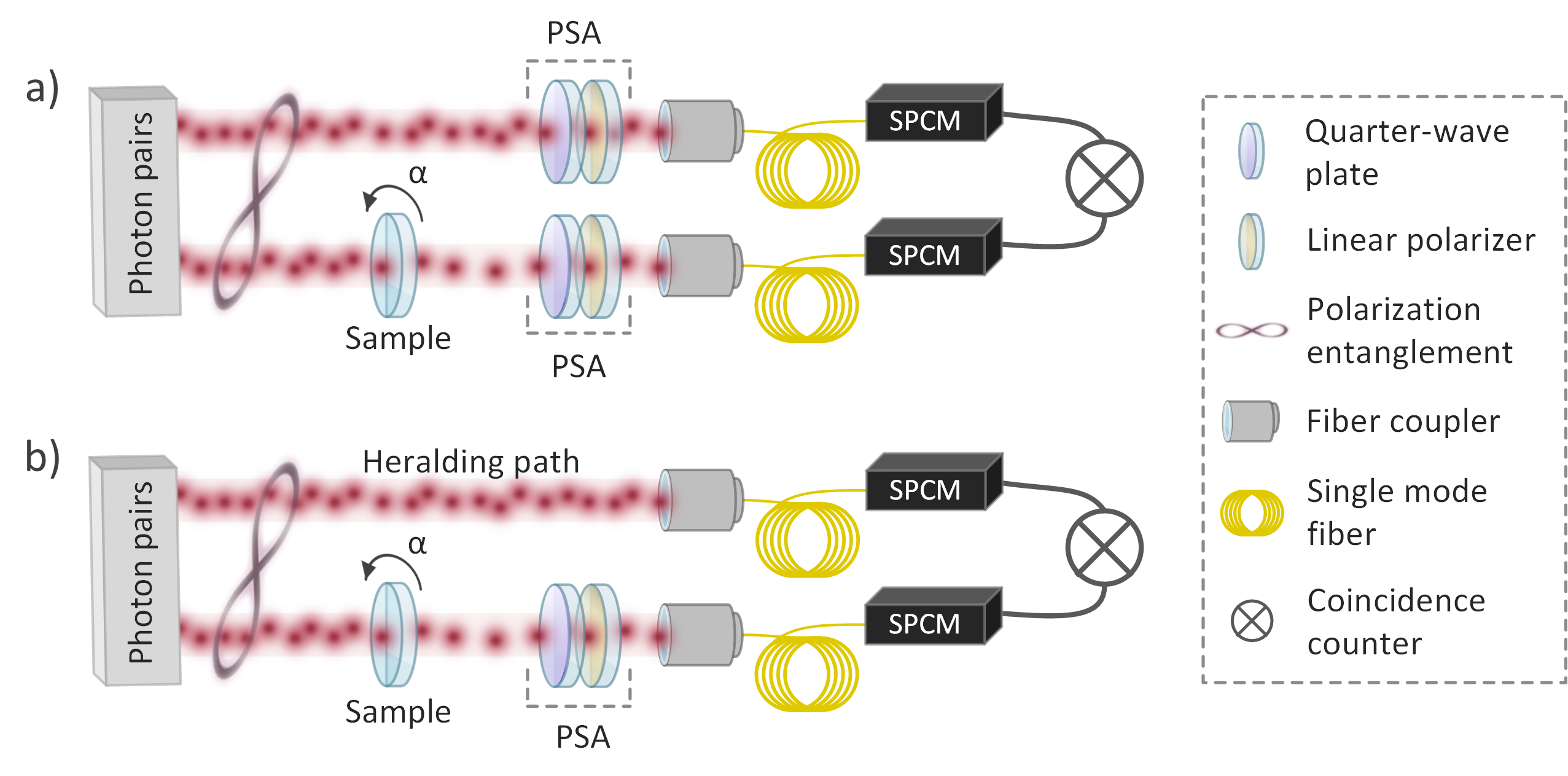} 
\caption{Optical arrangement used in the experiments. (a) Nonlocal quantum polarimetry using a pair of polarization-entangled photons (Bell state in the form of${1/\sqrt{2}}(\ket{HV}+\ket{VH}$). (b) Local quantum polarimetry. The partner photon from the entangled pair acts as a heralding one. PSA: polarization state analyzer; $\alpha$: angle of sample's principal optical axis orientation (fast axis for quarter-wave plate, transmission axis for linear polarizer); SPCM: single photon counting module.}
\label{fig:expScheme}
\end{figure}

The experimental setup that we used within the reported studies is outlined in~\cref{fig:expScheme}. A polarization Mach-Zehnder interferometer with two periodically-poled KTP (KTiOPO$_4$) crystals serves as the source of polarization-entangled photon pairs~\cite{Horn:2019}. The wavelength-degenerate photons at 810 nm are produced through spontaneous parametric down-conversion (SPDC) in type-II phase matching configuration. At the exit of the interferometer the photons are split into two spatially separated channels. In one of the channels we employ a combination of quarter- and half-wave plates to ensure the Bell state in the form of $\frac{1}{\sqrt{2}}(|HV\rangle+|VH\rangle)$. The preparation of the entangled photon pairs is not detailed in the figure. 

The performance of the source is evaluated via QST \cite{James:2001} using one detector per channel. For projective polarization detection we introduce rotating QWP and LP in both channels and fiber-coupled single photon counting modules. By using a time tagging device we acquire the coincidence events which are then used for reconstructing the density matrix by the maximum likelihood method \cite{Altepeter:2004}. The experimentally achieved state reaches a concurrence of 0.94 and a fidelity with the designed Bell state of 0.96.

During the sample evaluation, a QWP or a LP is introduced into the sample channel at several orientations. The angles between the optical axis of the sample (fast axis of the QWP, transmission axis of the LP) and the vertical axis in the laboratory frame were 0, 45$^{\circ}$, 90$^{\circ}$, and a randomly generated angled of 37$^{\circ}$. Following the concept from~\cref{fig:setup} we performed measurements using the one- and two-channel configurations. For the latter, the same procedure as for source performance evaluation was used, but with a sample influencing the polarization of one of the photons from the pair.

For the one-photon measurement we utilized the same source but polarization sensing was performed only in the sample channel. Taking into account that the number of coincidence events is much less than the single photon counts, the coincidence based measurement in two-channel configuration can not be directly compared with the single-photon experiment by simply using the same integration time. To overcome this issue we performed heralded single-channel measurements. For this, the QWP and LP in the reference arm were removed and the projective measurements were performed only in the sample channel. Also here, we counted the coincidence events between two channels, now the reference photon acting as a heralding one, and used these counts for one-photon density matrix reconstruction~\cite{Altepeter:2004}. Hereby we kept the number of photons contributing to the useful signal on the same level as in the two-channel measurement mode.

\begin{figure}[hb] 
\centering
\includegraphics[width=\textwidth]{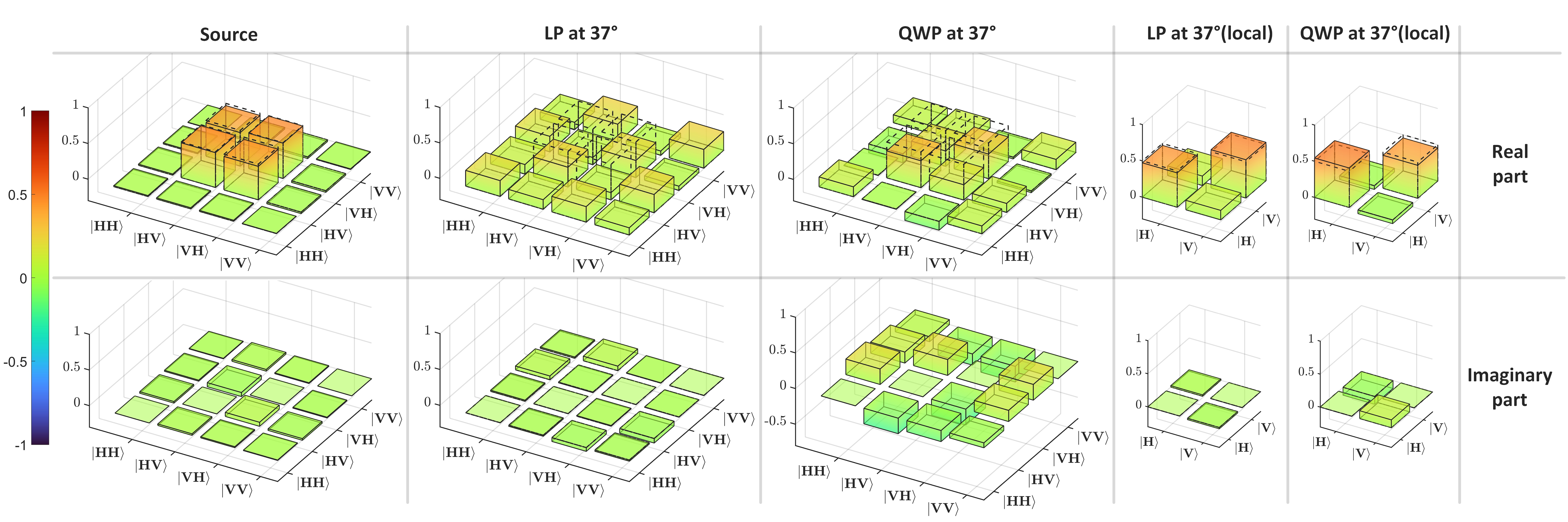} 
\caption{Representative density matrices of the probe state ("source"); two-photon state after the signal photon interacted with a linear polarizer (LP) and a quarter wave plate (QWP) oriented with their optical axis at 37$^{\circ}$ (averaged over 10 experimental runs); signal photon state after interaction with LP and QWP oriented at 37$^{\circ}$ (averaged over 10 experimental runs).}
\label{fig:expResults}
\end{figure}

The representative examples of the density matrix of the two-photon state entering the experimental arrangement, and outcomes from the two- and one-channel measurements are shown in~\cref{fig:expResults}. For both modes of the experiments we measured the density matrix of the quantum state altered by the sample, each repeated 10 times.


\section{ Theoretical Analysis of the Data  } \label{sec:Theoretical}

\begin{figure}
  \centering
  \includegraphics[width=18cm]{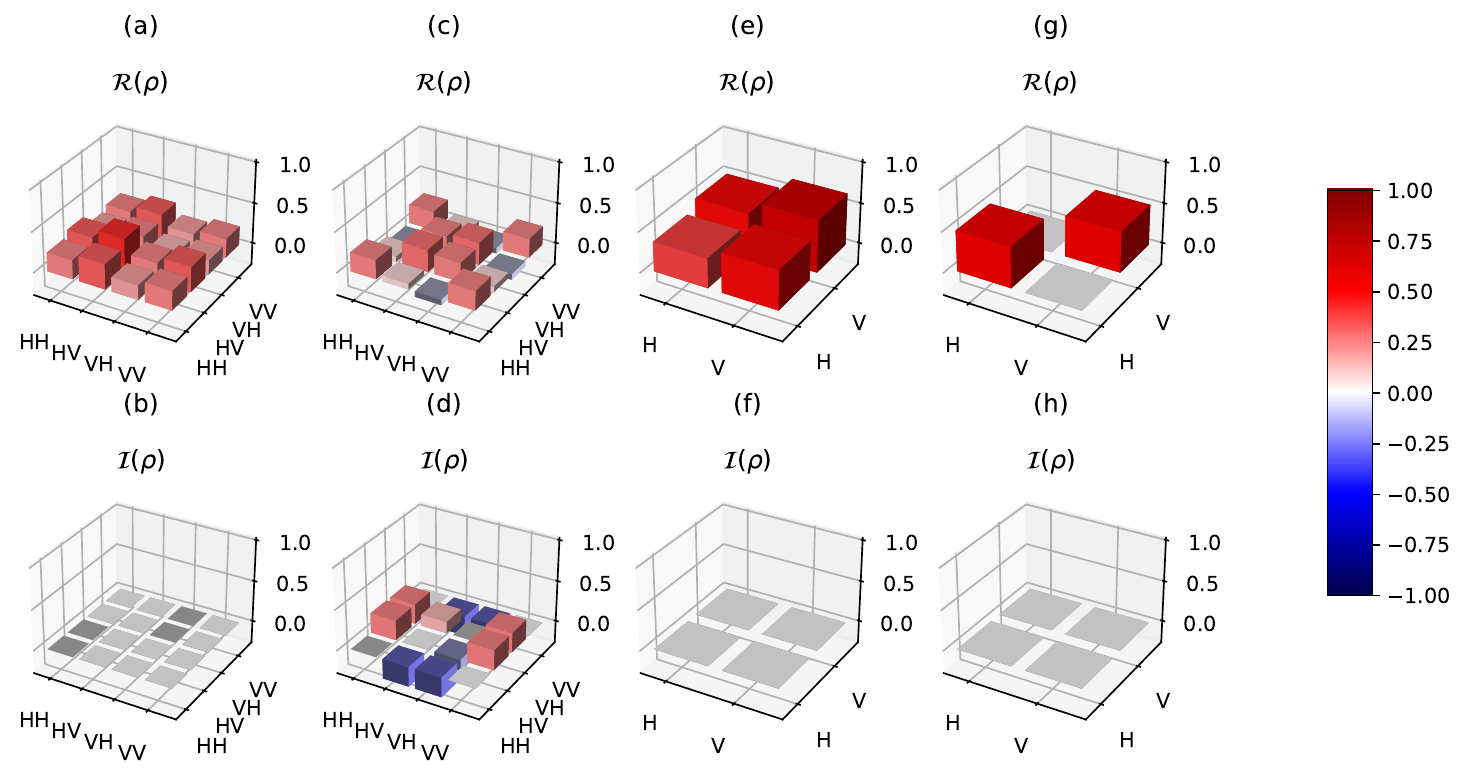}
  \caption{Theoretical results for the density matrix after each experiment shown as a histogram. Real (a) and imaginary (b) components of the density matrix for the non-local scheme with LP. Real (c) and imaginary (d) components of the density matrix for the non-local scheme with QWP. Real (e) and imaginary (f) components of the density matrix for the local scheme with LP. Real (g) and imaginary (h) components of the density matrix for the local scheme with QWP. Both LP and QWP are oriented at $37^{\circ}$.}\label{histogram_theory}
\end{figure}

In order to compare the experimental results given in~\cref{fig:expResults} with the theoretical predictions, in~\cref{histogram_theory} we calculated the expected output density matrices for each polarimetry task, which are shown in~\cref{histogram_theory}. We have used the open source package QuTiP for our calculations~\cite{JOHANSSON20131234}.

\begin{figure}
  \centering
  \includegraphics[width=10cm]{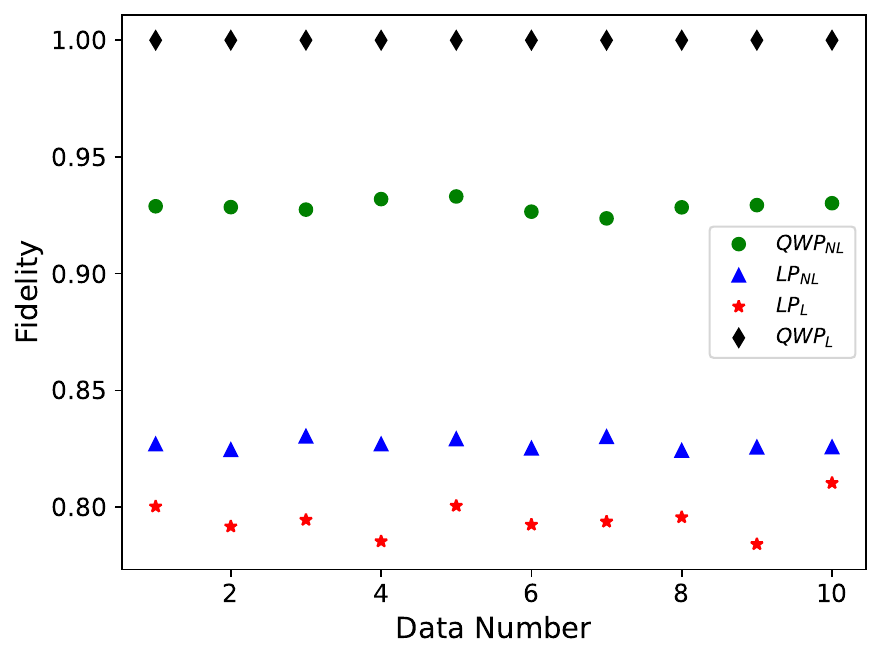}
  \caption{Fidelity between the tomographically reconstructed experimental density matrix and the theoretical density matrix for each experiment. Black diamonds, green circles, blue triangles and red stars represent the results for local QWP, non-local QWP, non-local LP and local LP respectively.}\label{fidelity_data}
\end{figure}

Upon comparing~\cref{fig:expResults} and~\cref{histogram_theory} we see that although they are qualitatively very close, the measured and calculated density matrix differ slightly. This is expected due to several sources of error and noise including state preparation error, ambient dark noise, imperfections in the optical elements, fluctuations of the photon source and errors in photon counting and post-processing. The errors are more pronounced for the local and non-local polarimetry experiments on LP. To gain a better understanding of the errors with respect to the theoretical predictions, in~\cref{fidelity_data} the fidelity of the experimental density matrices with their theoretical counterparts are given. Here we clearly see that the aforementioned experimental data are more distant to their theoretical counterparts. The goal in these experiments is to predict the value of $\alpha$ which we set to be equal to 37 degrees. We consider two different estimators to find the rotation angle of the optical devices.

From~\cref{out_2plp},~\cref{out_1plp} and~\cref{out_2pqwp} we can use the following expressions as estimators for $\alpha$.
\begin{equation}\label{est1}
\begin{aligned}
  \hat{\alpha}_1^{\text{LP}_{\text{NL}}}=\text{arctan}(\sqrt{\frac{\rho_{11}}{\rho_{22}}}),\\
  \hat{\alpha}_1^{\text{QWP}_{\text{NL}}}=\text{arccos}(\sqrt{2(\rho_{33}-\rho_{44})}),\\
  \hat{\alpha}_1^{\text{LP}_{\text{L}}}=\text{arctan}(\sqrt{\frac{\rho_{11}}{\rho_{22}}}).
\end{aligned}
\end{equation}

We can also define another estimator based on the value of angle that maximizes the fidelity between the experimental density matrix and the density matrix obtained by transforming the input state (entangled state for the non-local and maximally mixed state for the local) by that angle, using the corresponding transformation matrix related to the experiment
\begin{equation}\label{est2}
  \hat{\alpha}_2 =  \{\alpha^*\in[0,\pi/2]:\forall\alpha\in[0,\pi/2], \text{fidelity}(\rho_{\text{exp}},\varepsilon_{\alpha^*}(\rho_{\text{in}}))\geq \text{fidelity}(\rho_{\text{exp}},\varepsilon_{\alpha}(\rho_{\text{in}}))\}.
\end{equation}

\begin{figure}
  \centering
  \includegraphics[width=16cm]{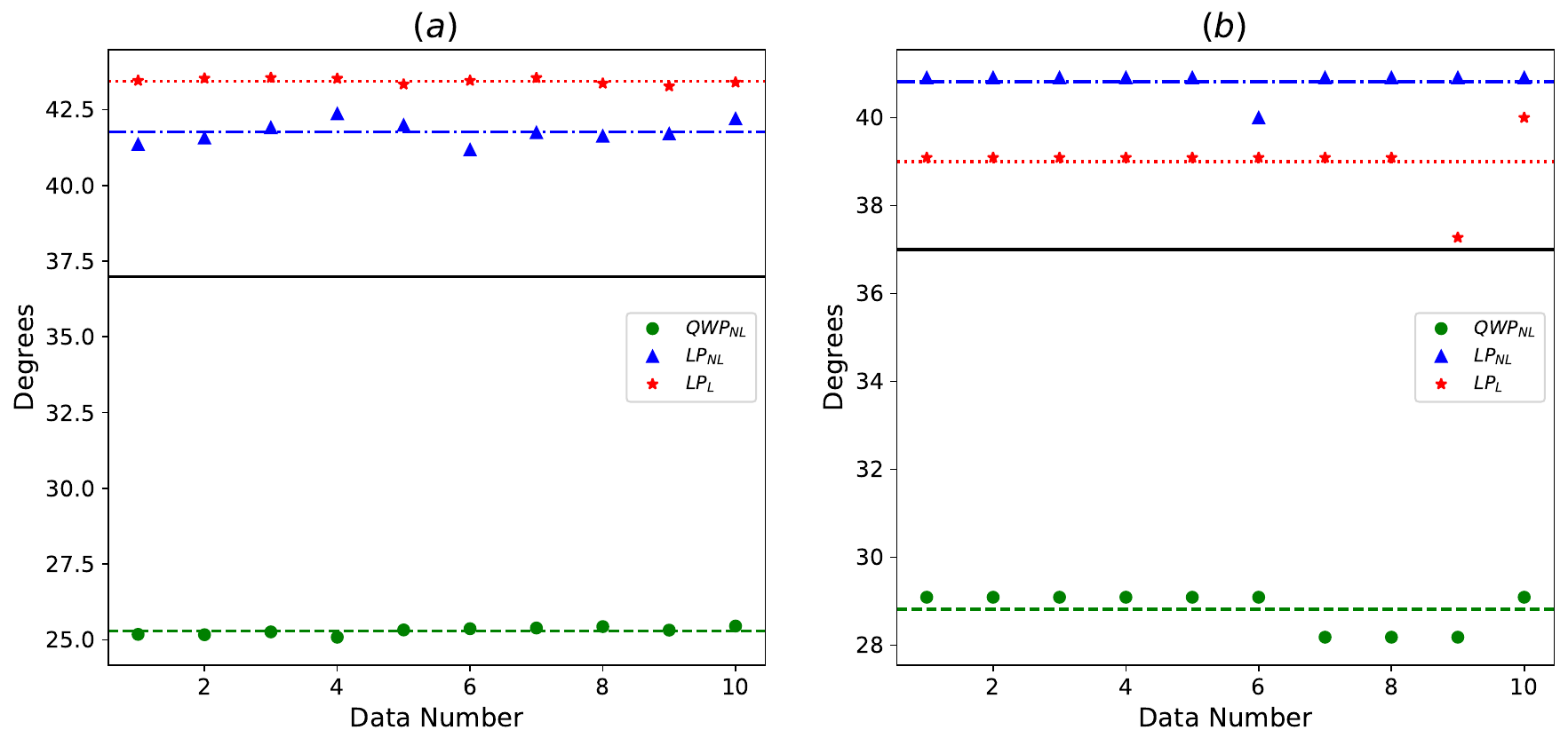}
  \caption{Estimated angle $\alpha$ for each experiment using two different estimators. (a) Estimated angles with $\hat{\alpha}_1$ using the trigonometric relations between the elements of the density matrices given in~\cref{est1}. The red stars and red dotted line are the angles estimated for local LP experiment and their average. The blue triangles and blue dot dashed line are the angles estimated for non-local LP experiment and their average. The green circles and green dashed line are the angles estimated for non-local QWP experiment and their averages. (b) Estimated angles with $\hat{\alpha}_2$ using the fidelity maximization procedure given in~\cref{est2}. The description of the data points and their averages are similar to part (a). The solid black line is the actual value of the angle at 37 degrees. The variances for the estimated angles in (a) are $\text{Var}(\hat{\alpha}_1^{\text{LP}_{\text{NL}}})=0.1202$, $\text{Var}(\hat{\alpha}_1^{\text{QWP}_{\text{NL}}})=0.0135$ and $\text{Var}(\hat{\alpha}_1^{\text{LP}_{\text{L}}})=0.0087$. The variances for the estimated angles in (b) are $\text{Var}(\hat{\alpha}_2^{\text{LP}_{\text{NL}}})=0.0744$, $\text{Var}(\hat{\alpha}_2^{\text{QWP}_{\text{NL}}})=0.1735$ and $\text{Var}(\hat{\alpha}_2^{\text{LP}_{\text{L}}})=0.4049$.}\label{angle_data}
\end{figure}

Here, $\varepsilon_{\alpha}$ represents a transformation operator using QWP or LP acting on the initial state and giving the final state and parameterized by $\alpha$. So all the estimators given in~\cref{est1} have their counterparts in~\cref{est2}. In~\cref{angle_data} the estimated angles given by these two sets of estimators are given.

Although theoretical predictions without considering any source of noise yield the correct result for the value of the angle, it is evident that the average value of the estimated angles for all polarimetric procedures and both estimators, does not equal to the real value of $\alpha=37$. Therefore, the results suggest that the angles estimated using experimental data contain statistical bias, reports for which are available in the literature~\cite{PhysRevA.108.042419,PhysRevLett.129.250503}. From~\cref{angle_data}(a) we see that indeed non-local polarimetry for LP using entangled photons gives a better estimate of $\alpha$ as compared to the local case. Also, we see that the QWP experiment has a far larger error compared with both of the LP experiments. However, even this is a considerable improvement compared to the local QWP experiment because in that case QFI is zero and no angle can be estimated. In~\cref{angle_data}(b) we see that applying the estimator defined by the fidelity maximization procedure, more accurate estimates for the values of $\alpha$ are obtained. In this estimator, unlike the one given by~\cref{est1}, the information from all of the elements of the density matrices are used and therefore generally it is expected that it will yield a better estimate. However, the effect of the noise on the estimation of the angle is highly non-trivial and comparing~\cref{angle_data}(a) and~\cref{angle_data}(b) we can see that although the accuracy of non-local polarimetry is higher than local case for the estimator $\hat{\alpha}_1$, the converse is true for $\hat{\alpha}_2$ . Additionally it must be stated that the estimate given in~\cref{est1} has one advantage over~\cref{est2} and that is the fact that one only needs two density matrix elements to come up with an angle estimate and full QST is not required.\\

Our results demonstrate that although one can use entanglement to enhance parameter estimation sensitivity, it is also important to consider the effect of realistic metrological experiments with various possible noise contributions on the accuracy of the estimated parameters. It is also important to design estimators which are more robust to loss of accuracy due to noise. For example, it is shown in~\cref{angle_data} that $\hat{\alpha}_1$ is more susceptible to the noise-induced bias compared with $\hat{\alpha}_2$. It is worth mentioning that for biased estimators the QCRB becomes~\cite{10.1116/1.5119961}
\begin{equation}\label{bias_qcrb}
  \text{Var}(\alpha)\geq \frac{(1+\partial_{\alpha}b)^2}{mF_Q},
\end{equation}
in which $b$ is the bias of the estimator. Our results show that, taking into account the various sources of experimental noise and uncertainty, estimated values for the angle will contain bias. To investigate the validity of this assertion and get more insight, in the next section we will perform QST on a set of numerically generated data which takes into account several sources of noise.


\section{ Theoretical Model of the Experimental noise } \label{sec:Conclusion}

However small it may be, no experiment is immune to noise. It is important to take into account the effects of noise on the results of the experiment, quantify it if it is possible, and take measures to mitigate such effects. In our experiment several sources of error exist that can potentially distort the data. In our theoretical analysis of the effect of noise we only consider four sources of noise, namely state preparation noise, dark counts, misalignment of measurement bases and Poisson noise in photon counting. Other possible sources of error such as imperfections of the photodetectors and other optical elements are neglected. We model the state preparation noise as
\begin{equation}\label{preparation_error}
  \hat{\rho}_{in} = q_1\hat{\rho} + (1-q_1)\hat{\rho}_r,
\end{equation}
in which $\hat{\rho}$ is the desired noiseless input state, $\hat{\rho}_r$ is a an average of a number of random density matrices and $q_1\in[0,1]$ is a real number quantifying the amount of preparation noise. In addition to the optical signal, the background noise can also trigger the photodetectors and induce dark counts. To model this effect, we assume that the quantum state of the background noise is given by a maximally mixed state and that it also contributes to the input of the photodetectors~\cite{Czerwinski_2022,Czerwinski_2021}.
\begin{equation}\label{dark_count}
  \hat{\rho}^{\prime} = q_2\hat{\rho}_{out} + (1-q_2)\frac{\hat{I}_d}{d}.
\end{equation}
Here, $q_2\in[0,1]$ is a real number quantifying the amount of dark counts and $\hat{I}_d$ is a $d\times d$ identity matrix. $\hat{\rho}_{out}$ is the density matrix at the output of LP or QWP considering the fact that the input state is $\hat{\rho}_{in}$. During the experiments, the measurement process is also subject to noise. In our calculations we model this noise by random rotations of the projection operators. The rotation operator is given by~\cite{Lohani_2020,Danaci_2021,Czerwinski_2022,Czerwinski_2021}.
\begin{equation}\label{random_rot}
  \hat{U}(\omega_1,\omega_2,\omega_3) = 
\begin{bmatrix}
  e^{i\omega_1/2}\text{cos}(\omega_3) & -ie^{i\omega_2/2}\text{sin}(\omega_3) \\
  -ie^{-i\omega_2/2}\text{sin}(\omega_3) & e^{-i\omega_1/2}\text{cos}(\omega_3), 
\end{bmatrix}
\end{equation}
in which $\omega_1$, $\omega_2$, $\omega_3$ are normally distributed parameters with zero mean and standard deviation $\sigma$. In this framework, $\sigma$ determines the amount of noise due to random rotations and misalignment in the measurement bases. The rotated projection operator becomes
\begin{equation}\label{proj_rot}
  \hat{\Pi}_i^{\prime} = (\hat{I}\otimes\hat{U})\hat{\Pi}_i(\hat{I}\otimes\hat{U}^{\dagger}).
\end{equation}
The coincidence counts, including the sources of noise, is given by
\begin{equation}\label{error_coin_count}
  n_{i}^{\prime} = \mathcal{N}_i\text{Tr}[\hat{\Pi}_i^{\prime}\hat{\rho}^{\prime}],
\end{equation}
in which $\mathcal{N}_i\in \text{Pois}(\mathcal{N})$ is randomly generated using the Poisson distribution with the mean value of $\mathcal{N}$. Using~\cref{error_coin_count} we include the effect of Poisson noise in our calculations for photon counting.

\begin{figure}
  \centering
  \includegraphics[width=16cm]{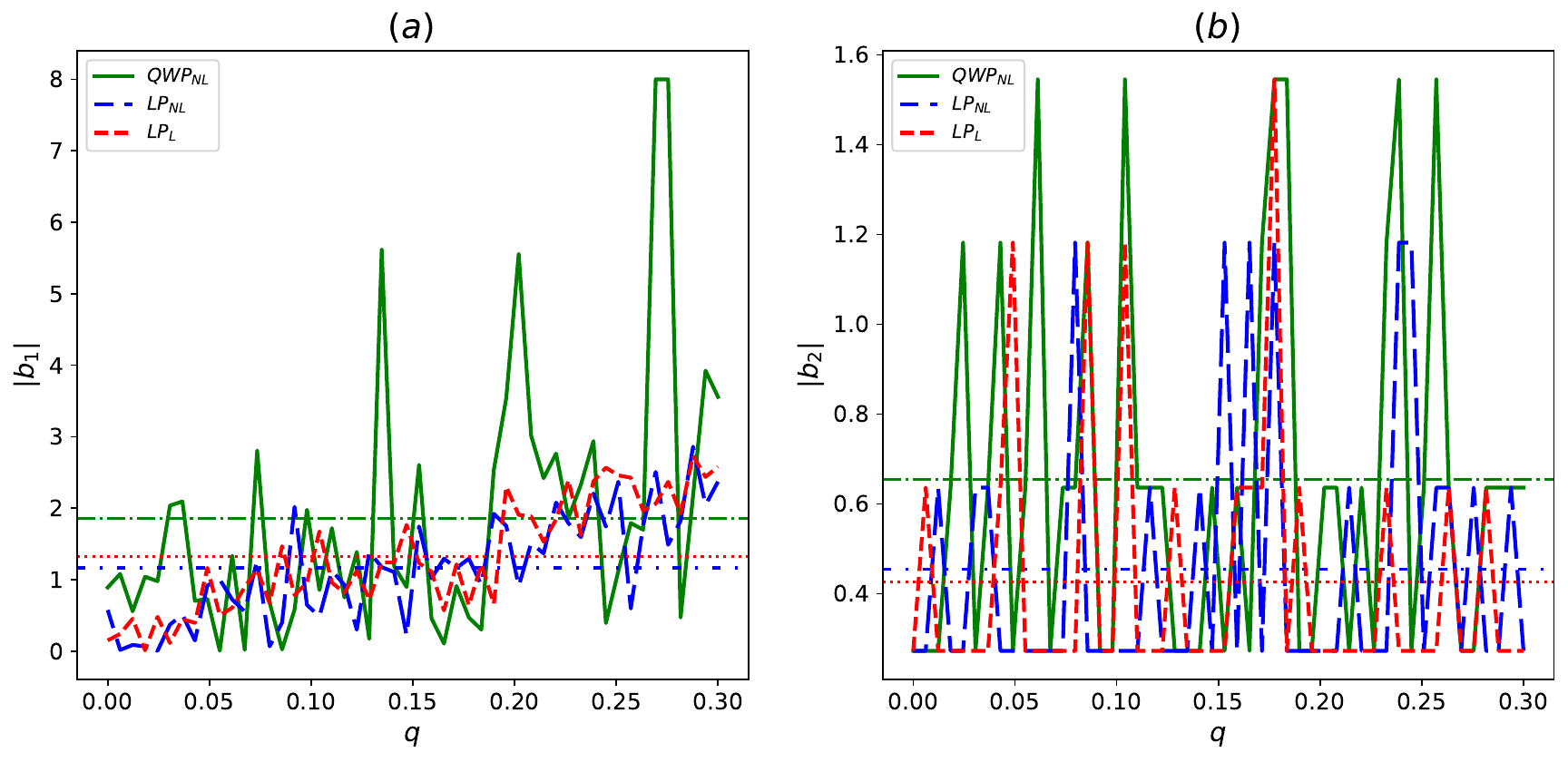}
  \caption{The magnitude of bias for the estimator (a) $\hat{\alpha}_1$ and $\hat{\alpha}_2$ (b) . In both (a) and (b) the solid green, long-dashed blue and the dashed red lines are $\hat{\alpha}_i^{\text{QWP}_{\text{NL}}}$, $\hat{\alpha}_i^{\text{LP}_{\text{NL}}}$ and $\hat{\alpha}_i^{\text{LP}_{\text{L}}}$ respectively. The expected values for magnitude of bias are shown by the dash-dotted green line for $\hat{\alpha}_i^{\text{QWP}_{\text{NL}}}$, loosely dash-dotted blue line for $\hat{\alpha}_i^{\text{LP}_{\text{NL}}}$, and dooted red line for $\hat{\alpha}_i^{\text{LP}_{\text{L}}}$.}\label{error_bias}
\end{figure}

Our goal is to show that the mentioned experimental noise models can induce bias in our parameter estimation problem. We don't intend to optimize the parameters of our noise model to quantify the experimental errors more accurately. For our calculations we take $\sigma=\pi/720$, $\mathcal{N}=5000$. The random density matrix in~\cref{preparation_error} is averaged out of 20 random density matrices. Also, to make presentation of our results simpler, we take $q_1=q_2=q$. The theoretical results for noisy QST for our chosen set of parameters is given in~\cref{error_bias}. Comparing~\cref{error_bias}(a) and~\cref{error_bias}(b) it is clear that compared to $\hat{\alpha}_1$ using $\hat{\alpha}_2$ as an estimator will result in a smaller bias. Our calculations show that for $\hat{\alpha}_1$, although there are fluctuations due to Poisson noise, the bias increases linearly with parameter $q$ as shown in~\cref{error_bias}(a). For our selected set of parameters we see that in~\cref{error_bias}(a) that in case of using $\hat{\alpha}_1$ as the estimator $\hat{\alpha}_1^{\text{LP}_{\text{NL}}}$ results in the smallest bias. However, using the estimator $\hat{\alpha}_2$, as shown in~\cref{error_bias}(b), $\hat{\alpha}_2^{\text{LP}_{\text{L}}}$ yields the smallest value for bias. 
Our results clearly show that including sources of error in our calculations will induce bias in the estimated angles. Therefore, to fully utilize the metrological quantum advantage one can either try to reduce the effect of noise on the experimental results, or design estimators that are less susceptible to noise.


\section{ Conclusion  } \label{sec:Conclusion}


Our theoretical and experimental framework illustrates that employing nonlocal polarimetry with an initially entangled state leads to improved precision and accuracy in extracting information about sample characteristics compared to the case of local polarimetry. Our results show that using quantum entanglement can increase the precision and accuracy depending on the chosen estimators and existing noise channels. We have seen that the experimental noise induces bias in our estimators and therefore reduces the accuracy of the estimated results. Therefore, to harness the full potential of quantum states for metrological tasks, one needs to address the issue of noise-induced bias as well.

Our investigation reveals the presence of a quantum nonlocality advantage within the ambit of two-channel entangled state quantum polarimetry. However, it is pertinent to note that this quantum advantage remains suboptimal. Within the realm of quantum metrology, the utilization of NOON states, expressed in the form $(\ket{HH}+\ket{VV})/\sqrt{2}$, could conceivably offer a higher enhancement in precision compared to local polarimetry~\cite{6999929}. Furthermore, such states could potentially be harnessed to attain quadratic advantages with increasing photon numbers. While our current study identifies the nonlocal advantage, the pursuit of optimization for this nonlocal scheme remains an open challenge, awaiting exploration in future investigations. 

\acknowledgments
This work has been funded by the European Union’s Horizon 2020 research and innovation programme (Grant Agreement No. 899580); the German Federal Ministry of Education and Research (FKZ 13N14877 and 13N15956); and the Cluster of Excellence “Balance of the Microverse” (EXC 2051-Project No. 390713860). V.B. thanks for funding of this work also through the ProChance-career program of the Friedrich Schiller University Jena. A. P. thanks Aaron Z. Goldberg from National Research Council of Canada for fruitful discussions.

%

\end{document}